\documentclass[preprint]{revtex4}  
\usepackage{amsfonts}
\usepackage{amsmath}
\usepackage{cancel}
\usepackage{graphicx}
\usepackage{mathcomp}
\usepackage{color}

\begin{document}

\title[Algebraic Sheath Model]{An algebraic RF sheath model for all excitation waveforms \\ and amplitudes, 
                               and all levels of collisionality}

\author{A.E.~Elgendy, H.~Hatefinia, T.~Hemke, M.~Shihab, A.~Wollny, D.~Eremin, T.~Mussenbrock, and R.P.~Brinkmann}

\address{Ruhr-University Bochum \\ Institute for Theoretical Electrical Engineering\\ D-44780 Bochum, Germany}

\date{\today}

\begin{abstract}
The boundary sheath of a low temperature plasma comprises typically only a small fraction of its volume but is
responsible for many aspects of the macroscopic behavior.
 A thorough understanding \linebreak of the sheath dynamics is therefore of theoretical and practical importance. 
This work focusses on \linebreak the so-called ``algebraic'' approach which strives to describe the electrical behavior of RF modulated boundary 
sheaths in closed analytical form,
i.e., without the need to solve differential equations. \linebreak
A mathematically simple, analytical expression for the charge-voltage relation of a sheath is presented which holds for all excitation wave forms and amplitudes and 
covers all regimes from the collision-less motion at low gas pressure to the collision dominated motion at gas high pressure. \linebreak
A comparison with the results of self-consistent particle-in-cell simulations is also presented.
\end{abstract}

\maketitle

\section{Introduction}

To study the plasma boundary sheath with algebraic models has a long tradition 
\cite{Lieberman1988,Lieberman1989,MetzeErnieOskam1986,SheridanGorree1989,HorwitzPuzzer1990,RobicheBoyleTurnerEllingboe2003,JiangMaoWang2006,DewanMcNallyHerbert2001,
DewanMcNallyHerbert2002,TurnerChabert2012}. \linebreak
In contrast to more complex models which require time-consuming computer simulations, \linebreak
algebraic sheath models execute effectively in zero time. This makes them suitable for many 
practical purposes, for example for the model based real-time control of plasma processes. \linebreak
Of course, the validity of simplified descriptions is always an issue. Algebraic sheath models should thus be carefully derived
from first principles and thoroughly tested
against more complex (and more physical) approaches such as \textit{particle in cell} (PIC) simulations.

In the present manuscript we propose a novel algebraic model for the electrical behavior of the RF modulated plasma boundary sheath. 
Our investigation is motivated by a critical assessment of the pioneering (and still ``classical'') algebraic sheath models 
which were proposed by Lieberman twenty-five years ago \cite{Lieberman1988,Lieberman1989}. In many aspects, our approach is similar:\linebreak
We focus on the RF regime, where the applied radio frequency lies between the plasma frequencies  of ions and electrons,
$\omega_{\rm pi} \ll \omega_{\rm RF} \ll \omega_{\rm pe}$,  consider only
one species of singly charged positive ions without any ``chemistry'', and assume a one-dimensional Cartesian geometry.\linebreak
We endeavor, however, to correct the three fundamental weaknesses of the Lieberman models,
namely that they are limited to the case of a single driving frequency, to the regime of large applied voltages 
(compared to the thermal voltage $T_{\rm e}/e$, with $T_{\rm e}$ the electron temperature), and to the two limiting cases of either highly collisional or completely 
collision-free motion. In other words, our goal is an algebraic model which captures the plasma sheath
behavior in a wide range of frequencies, waveforms, amplitudes, and collisionality.

Our manuscript is organized as follows: In the next section, we describe our starting point, the ``standard model'' 
of the RF modulated sheath. After reviewing the Lieberman approach, \linebreak we return to the standard model and employ the  
\textit{advanced algebraic approximation} \cite{Brinkmann2007,Brinkmann2009,Brinkmann2011} to transform it -- practically without any loss of accuracy -- into a mathematically simpler form. \linebreak
The outcome of this first step is a valid  sheath model of its own, but not yet of a closed form.
 We thus take a second step, employing a sequence of additional, more drastic approximations. 
(A two-step approach is chosen because the intermediate model is better suited to assess the   effects of the 
``drastic appproximations'' than the original one.)  
The result of the second step is our novel algebraic model, which then will be thoroughly tested against PIC simulations. Some conclusions and final remarks 
are given in the last section.

\section{Mathematical model}

\begin{figure}[h!]
  \centering
 \includegraphics[width=0.8\textwidth]{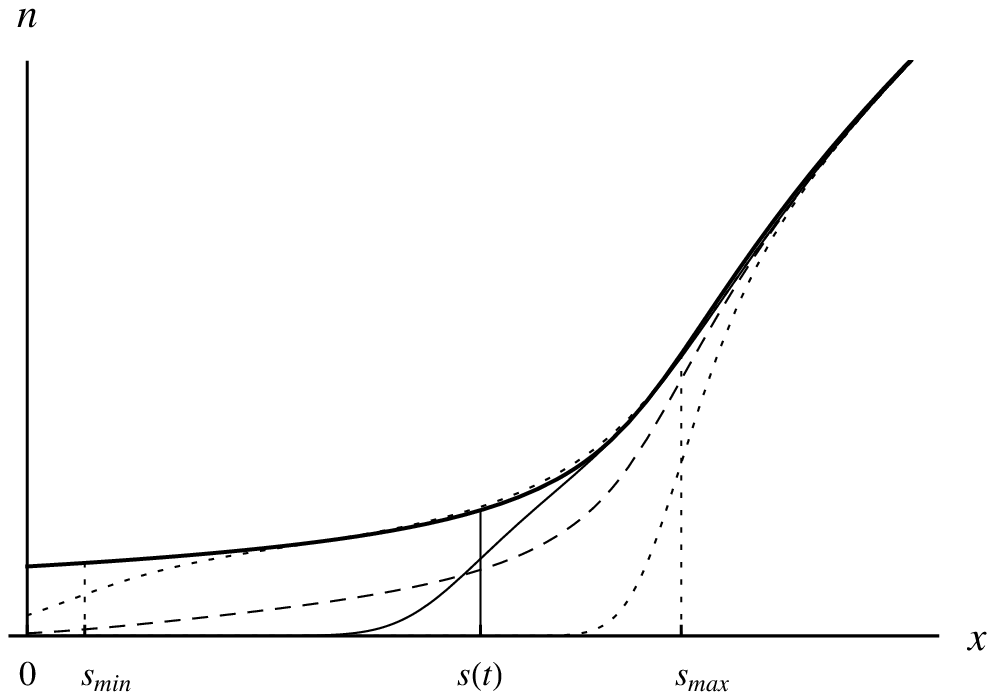}
 \caption{Schematic sketch of the particle densities in an RF driven sheath. The $x$-axis points from the electrode $x_{\rm E}=0$ into the bulk. (Note that several other conventions are used in the literature.)
Shown are the stationary ion density $n_{\rm i}(x)$ (solid) and the momentaneous electron density $n_{\rm e}(x,t)$ at a certain RF phase $t$ (thin). The average electron 
         density ${\bar n}_{\rm e}(x)$ is dashed. 
         Also shown are the momentaneous location $s(t)$ of
         the equivalent electron edge, and the minimal and
         maximal values of that quantity, $s_{\rm min}$ and $s_{\rm max}$.
         Note that the value of $s_{\rm min}$ is generally different from the position of the
         electrode, and that $s_{\rm max}$ is only an approximate indicator of the sheath edge.}\label{FIG1SchematicPlot}
 \end{figure}

For the description of the sheath, we employ what may be called the ``standard model''. \linebreak
It assumes the RF regime, i.e., that the exciting radio frequency $\omega_{\rm RF}$
lies between the plasma frequencies of  ions and  electrons, 
$\omega_{\rm pi} \ll \omega_{\rm RF} \ll \omega_{\rm pe}$, and also imposes the 
length scale ordering $\lambda_{\rm D} \ll s  \ll L \approx \lambda_{\rm ion}$,
where $\lambda_{\rm D}$ is the Debye length, $s$ the sheath thickness, $L$~the system length,
and $\lambda_{\rm ion}$ the ionization length scale. The geometry is one-dimensional; we consider the spatial interval $[x_{\rm E}, x_{\rm B}]$,
where $x_{\rm E}=0$ denotes the location of the electrode and $x_{\rm B}$ a point far enough into the 
bulk so that quasineutrality prevails for all phase points.\linebreak Fig.~\ref{FIG1SchematicPlot} shows the coordinates and other conventions.
Symbols have their standard meaning. \linebreak
We also define the phase interval $[0,T]$, and the corresponding phase average as
\begin{align}
	 \bar{f} = \frac{1}{T}\int_0^T f(t)\, dt. \label{PhaseAverage}
\end{align}

The electron part of the model consists of the equation of continuity with ionization and recombination neglected and 
of the relation of Boltzmann equilibrium,
\begin{align}
 &\frac{\partial n_{\rm e}}{\partial t} + \frac{\partial}{\partial x}\left( n_{\rm e} v_{\rm e}\right) = 0, 
\label{}
\\[1.0ex]
 & T_{\rm e} \frac{\partial n_{\rm e}}{\partial x} + e n_{\rm e} E = 0. \label{BoltzmannEquilibrium}
\end{align}
Ions are assumed to experience no modulation. Their equation of continuity can be integrated to express the constancy 
of the ion flux flowing in negative $x$-direction,
\begin{align}
	n_{\rm i} v_{\rm i} =-\Psi_{\rm i} = \rm const. \label{IonContinuity}
\end{align}
The equation of motion describes the acceleration of ions under the action of the phase-averaged electrical field $\bar E$ 
and the friction due to collisions with the neutral background.\linebreak  In sheath models, the latter term is typically modeled by the assumption of 
a constant ion mean free path $\lambda_{\rm i}$ which is valid for strong electrical fields. (A more general ansatz for the friction gives
qualitatively similar results but with less transparent formulas).
\begin{align}
   v_{\rm i} \frac{\partial v_{\rm i}}{\partial x} = \frac{e}{m_{\rm i}} \bar{E} - \frac{\pi|v_{\rm i}|}{2\lambda_{\rm i}} \, v_{\rm i}.
	\label{IonEquationOfMotion}
\end{align}
The field is described by Poisson's equation,
\begin{align}
	 \epsilon_0 \frac{\partial E}{\partial x} = 		e \left(n_{\rm i} - n_{\rm e} \right).
	\label{PoissonEquation}
\end{align}

The model is mathematically completed by a set of boundary conditions and constraints. \linebreak Sheath-like solutions as shown in Fig.~\ref{FIG1SchematicPlot} 
are sought which obey asymptotic quasineutrality and transport equilibrium (drift regime) for $x\to x_{\rm B}$ and electron depletion for  $x\to x_{\rm E}$. \linebreak
(Together, these assumptions remove all degrees of freedom of the ODE system except one.)
The RF modulation of the sheath is introduced by prescribing the total current density $j(t)$, \linebreak which is divergence free and  a spatial constant in 1d. 
We express it as the sum of a temporally constant part $\bar j$ and a periodic, average free, not necessarily harmonic part $\tilde{j}(t)$,
\begin{align}
	  \epsilon_0 \frac{\partial E}{\partial t} + j_{\rm e} - e \Psi_{\rm i} = j(t) = \bar{j} + \tilde{j}(t).
\end{align}
This relation can be cast in a more explicit form. We define the sheath charge per area $Q(t)$ as the integral of the 
charge density between the electrode $x_{\rm E}$ and the  bulk point $x_{\rm B}$, 
\begin{align}
	Q(t) = \int_{x_{\rm E}}^{x_{\rm B}} e (n_{\rm i} - n_{\rm e}) \, dx.
\end{align} 
Integrating Poisson's equation from $x_{\rm E}$ to $x_{\rm B}$ yields a relation between $Q(t)$ and the electrical field at the electrode;
provided that the field at $x_{\rm B}$ can be neglected:
\begin{align}
E (x_{\rm E})	 = -\frac{1}{\epsilon_0} Q(t) + E (x_{\rm B}) \approx -\frac{1}{\epsilon_0} Q(t). 
\end{align}
The electron current ${j}_{\rm e}$ at the electrode can be found by the Hertz-Langmuir formula which expresses the flux
as the product of the projected thermal speed and the local density \cite{Riemann}.\linebreak Taking into account an effective sticking factor $s_{\rm e}$
which may be, contrary to common belief, not equal to unity \cite{Bronold2012}, we set:
\begin{align}
  j_{\rm e}(x_{\rm E})  = s_{\rm e} e \sqrt{\frac{T_{\rm e}}{2\pi m_{\rm e}}} \, n_{\rm e}(x_{\rm E}).
\end{align}
Integrating the Boltzmann relation from $x_{\rm E}$ to $x_{\rm B}$, we can express the electron density at $x_{\rm E}$\linebreak
 in terms of presumably constant density at $x_{\rm B}$
and the sheath voltage $V_{\rm sh}$:
\begin{align}
	n_{\rm e}(x_{\rm E}) = 	n_{\rm e}(x_{\rm B}) \exp\left(-\frac{e V_{\rm sh}}{T_{\rm e}}\right)\!.
\end{align}
Here, the sheath voltage is defined as the field integral from $x_{\rm E}$ to $x_{\rm B}$, with the negative sign reflecting the fact that the field is oriented into the negative $x$-direction:
\begin{align}
	V_{\rm sh} = - \int_{x_{\rm E}}^{x_{\rm B}} E \, dx.
\end{align}
Together, these assumptions gives the current balance as follows, which is the justification for the popular diode model \cite{MetzeErnieOskam1986}
of the plasma boundary sheath (Fig.~2): The total current $j(t)$\linebreak  is the sum of a capacitive part (represented by a nonlinear capacitor),
 an exponential electron part (represented by a diode) and a constant part (represented by a current source):  
\begin{align}
	j(t) =  - \frac{dQ}{dt}  + s_{\rm e} e \sqrt{\frac{T_{\rm e}}{2\pi m_{\rm e}}} \, n_{\rm e}(x_{\rm B}) \exp\left(-\frac{e V_{\rm sh}}{T_{\rm e}}\right)  - e \Psi_{\rm i}. \label{DiodeModel}
\end{align}

Taking the phase average of the current balance relation \eqref{DiodeModel} yields a representation of the DC current characteristics of the sheath:
\begin{align}
	 \bar{j} =    s_{\rm e} e \sqrt{\frac{T_{\rm e}}{2\pi m_{\rm e}}} \, \bar{n}_{\rm e}(x_{\rm E})  - e \Psi_{\rm i}.
	\label{DCCurrent}
\end{align}
When the electrode is current-free or ``floating'', the DC current density $\bar{j}$ is equal to zero. 
To determine the corresponding sheath condition, one must solve
\begin{align}
	 s_{\rm e} e \sqrt{\frac{T_{\rm e}}{2\pi m_{\rm e}}} \, \bar{n}_{\rm e}(x_{\rm E})  = e \Psi_{\rm i}.
	\label{ElectrodeCondition}
\end{align}

In the fluctuating part of the current balance, we can  neglect the
(in~comparison with the RF current)  typically small electron current $\tilde{j}_{\rm e}$ and write
\begin{align}
	\tilde{j}(t) =  - \frac{dQ}{dt}.
\end{align}
It is advantageous to split $Q$ into an average $\bar{Q}$ and a fluctuating, average-free part $\tilde{Q}(t)$.\linebreak 
The average sheath charge defines the mean sheath thickness $\bar{s}$ via
\begin{align}
		\bar{Q} = \frac{1}{T} \int_0^T Q(t)\, dt =: \int_{x_{\rm E}}^{\bar{s}}  e n_{\rm i}(x) \, dx. 
		\label{AverageCharge}
\end{align}
For the fluctuating sheath charge, we get the explicite representation
\begin{align}
	\tilde{Q}(t) = \int_0^t \tilde{j}(t^\prime) \, dt^\prime+\frac{1}{T}\int_0^T  t^\prime \, \tilde{j}(t^\prime) \, dt^\prime.
\end{align}
We take the view that the fluctuating charge $\tilde{Q}(t)$ is the control parameter of the modulation \linebreak 
and define $\tilde{Q}_{\rm min}$ and $\tilde{Q}_{\rm max}$ as its minimum and maximum values within the phase cycle $[0,T]$. 
For vanishing modulation, both are zero, otherwise $\tilde{Q}_{\rm min}$ is negative and $\tilde{Q}_{\rm max}$ is positive.
Their absolute values are different unless $\tilde{Q}(t)$ is symmetric. As additional characterization  we introduce
 the effective modulation amplitude
\begin{align}
\Delta Q = \sqrt{\frac{1}{T} \int_0^T 	{\tilde Q}(t)^2\, dt} \label{EffectiveModulationAmplitude}.
\end{align} 

\section{Dimensionless notation and characteristic numbers}

It is advantageous to write the model dimensionless. 
However, some care must be taken, as an unfortunate choice of units may obscure the scaling relations.
We take as basis the \linebreak voltage scale $\hat{V}$ of the sheath; for weak modulation it is several $T_{\rm e}/e$,
for strong modulation it is equal to the applied voltage. The other units are calculated using collisionless relations. \linebreak
Altogether, we make the following substitutions, with the prime (which is dropped soon)\linebreak denoting dimensionless quantities:
\begin{align}
   &x \to \hat{s} x^\prime = ({\hat{V}^3 \epsilon_0^2}/{e m_{\rm i} \Psi_{\rm i}^2})^{1/4}	\, x^\prime,\\ 
	 &t \to \omega_{\rm RF}^{-1}\, t^\prime = 2\pi T t^\prime,\\ 
	 &n \to \hat{n} n^\prime = ({m_{\rm i}  \Psi_{\rm i}^2}/{e \hat{V}})^{1/2} \, n^\prime,\\ 
	 &v_{\rm i} \to \hat{v} v_{\rm i}^\prime = ({e \hat{V}}/{m_{\rm i}})^{1/2}\,v_{\rm i}^\prime,\\ 
   &Q \to \hat{Q} Q^\prime =(e m_{\rm i} \hat{V} \Psi ^2 \epsilon_0^2)^{1/4} Q^\prime, \\ 
   &\tilde{j} \to \hat{j} \tilde{j}^\prime = \omega_{\rm RF} (e m_{\rm i} \hat{V} \Psi ^2 \epsilon_0^2)^{1/4}  \tilde{j}^\prime, \\ 
   &\bar{j} \to e \Psi_{\rm i}\, \bar{j}^\prime, \\ 
	 & E \to \hat{E} E^\prime =  ({e\hat{V} m_{\rm i} \Psi_{\rm i}^2}/{\epsilon_0^2})^{1/4}E^\prime .
\end{align}
We also introduce dimensionless numbers (and their typical values). The ratio of the length scale to the ion mean free path,
with $\pi/2$ absorbed, is the collisionality $\nu$,
\begin{align}
	 \nu  =\frac{\pi}{2}\frac{\hat{s}}{\lambda_{\rm i}} = \frac{\pi}{2}\frac{ ({\hat{V}^3 \epsilon_0^2}/{e m_{\rm i} \Psi_{\rm i}^2})^{1/4}}
	{\lambda_{\rm i}} 		\approx 0.01 \ldots 100,
\end{align}
the ratio of the electron voltage to the voltage scale is the thermal parameter $\vartheta$,
\begin{align}
	\vartheta = \frac{T_{\rm e}}{e \hat{V}}  \approx 0.01 \ldots 0.2,
\end{align}
and the combination of the  electron sticking factor $s_{\rm e}$ and the square root of the mass ratio,
with  $2\pi$ absorbed, is the effective sticking parameter $\sigma$,
\begin{align}
	\sigma = s_{\rm e}  \sqrt{\frac{  m_{\rm i}   }{2\pi m_{\rm e}}} \approx 10 \ldots 300
\end{align}
Another parameter, the ratio of the RF current to the DC current, does not appear explicitly,
but it should be noted that it is typically large (except for non-modulated sheaths):
\begin{align}
\frac{ \hat{j}}{ e \Psi_{\rm i}}= \frac{\omega_{\rm RF}\hat{Q}}{e \Psi_{\rm i}}\approx 30 \ldots 300.
\end{align}

\pagebreak

The dimensionless equations then consist of several subgroups: The stationary ion model  involves the phase-averaged field $\bar E$,
\begin{align}
&	n_{\rm i} v_{\rm i} =-1, \label{DimensionlessIonContinuity} \\
&	   v_{\rm i} \frac{\partial v_{\rm i}}{\partial x} = \bar{E} - \nu  |v_{\rm i}| \, v_{\rm i}.
	\label{DimensionlessIonEquationOfMotion}	
\end{align}
The electron model consists of Boltzmann's equilibrium and Poisson's equation, 
\begin{align}
& \vartheta\,\frac{\partial n_{\rm e}}{\partial x} +  n_{\rm e} E   = 0, \label{DimensionlessBoltzmannEquilibrium}\\
& \frac{\partial E}{\partial x} = 	\left(n_{\rm i} - n_{\rm e} \right).
	\label{DimensionlessPoissonEquation}
\end{align}
It is parametrical modulated by the condition that the fluctuating sheath charge follows a given time function
$\tilde{Q}(t)$ which is related to the fluctuating RF current $ \tilde{j}(t)$,
\begin{align}
	\int_{\bar{s}}^{x_{\rm B}} n_{\rm i} - n_{\rm e} \, dx \stackrel{!}{=} \tilde{Q}(t)=  \int_0^t \tilde{j}(t^\prime) \, dt^\prime+\frac{1}{2\pi}\int_0^{2\pi}  t^\prime \, \tilde{j}(t^\prime) \, dt^\prime.
\end{align}
The position of the electrode $x_{\rm E}$  relative to the solution is determined by chosing a particular point of the 
DC (phase-averaged) current voltage curve 
\begin{align}
	\bar{j}   =    \sigma  \sqrt{\vartheta} \,\, \bar{n}_{\rm e}(x_{\rm E})  - 1 = \sigma  \sqrt{\vartheta} \, n_{\rm e}(x_{\rm B})\,
	\overline{\exp\left(-\frac{ V_{\rm sh}}{\vartheta }\right)}  - 1, \label{DimensionlessCurrentVoltageRelation}
\end{align}
where the sheath voltage is calculated as 
\begin{align}
	V_{\rm sh} = - \int_{x_{\rm E}}^{x_{\rm B}} E \, dx. \label{DimensionlessSheathVoltage}
\end{align}
For the particularly important case of a floating electrode or wall, the current relation reduces to
the condition for a current-free sheath,
\begin{align}
	\sigma  \sqrt{\vartheta} \, \bar{n}_{\rm e}(x_{\rm E})=\sigma  \sqrt{\vartheta} \, n_{\rm e}(x_{\rm B})\, \overline{\exp\left(-\frac{ V_{\rm sh}}{\vartheta }\right)} 
	= 1. \label{DimensionlessFloatingRelation}
\end{align}
Once the electrode position is known, also the average sheath charge can be calculated:
\begin{align}
		\bar{Q} = \frac{1}{2\pi} \int_0^{2\pi} Q(t)\, dt =: \int_{x_{\rm E}}^{\bar{s}}  n_{\rm i}(x) \, dx. 
		\label{DimensionlessAverageCharge}
\end{align}

The problem posed by the listed equations is quite intricate. The two subsystems alone \linebreak are nonlinear differential equations, 
but they are coupled by the operation of phase averaging.\linebreak Altogether, the sheath problem thus amounts to a system of nonlinear 
integro-differential equations 
for which no analytical solutions are known. Of course, numerical solutions of the system can be easily constructed, but this  misses the point of an 
analytical treatment. Clearly, a ``short-cut'' is needed which allows to construct explicit expressions for the phase-averaged
quantities $\bar{E}$ or $\bar{n}_{\rm e}$. (See Fig.~\ref{Shortcut}).

\section{The Lieberman approach and its limitations}

In his pioneering work, Lieberman proposed such a shortcut \cite{Lieberman1988,Lieberman1989}. His approach consisted of three
approximations and specializations:
First, he imposed Godyak's step model \cite{GodyakGhanna1979}. This approximation rests on the fact that the thermal electron 
voltage $T_{\rm e}/e$ is small compared to the applied sheath voltage $V_{\rm sh}$; i.e., in our notation, that the thermal parameter $\vartheta$ is small.
It replaces the Boltzmann relation for the electrons by the assumption that the 
density of the electrons is zero below the electron edge $s(t)$ and equal to the ion density above:  
\begin{align}
&n_{\rm e}(x,t)  =  \left\{
\begin{array}{c@{\quad:\quad}l}
            n_{\rm i}(x)  & s < s(t), \\
            0 & x \ge s(t).
          \end{array}
\label{StepModelDensity}
\right.
\end{align}
The value of $s(t)$ is related to the sheath charge $Q(t)$;
\begin{align}
	Q(t) = \int_{x_{\rm E}}^{s(t)} n_{\rm i}(x)  \, dx.
\end{align} 
Substituted into Poisson's equation, the step model allows to calculate the electrical field, where the integration constant is chosen
so that the field vanishes at the step:
\begin{align}
E(x,t)  =  \left\{
\begin{array}{c@{\quad:\quad}l}
            \displaystyle\int_{x}^{s(t)}n_{\rm i}(x)  & s < s(t), \\
            0 & x \ge s(t).
          \end{array}
\label{StepModelField}
\right.
\end{align}
Lieberman's second simplification was that he did not study the full range of collisionality but only the limiting cases of completely collisional and
completely collisionless dynamics.\linebreak  In our notation, they are given by the ion equations of motion
\eqref{CollisionlessDimensionlessIonEquationOfMotion} or \eqref{CollisionalDimensionlessIonEquationOfMotion}
 instead of \eqref{DimensionlessIonEquationOfMotion}.\linebreak Both assumption enable the explicit evaluation of the ion model by establishing 
an algebraic relation between the ion speed and either the averaged  potential $\bar{\Phi}$ (in the collisionless case)\linebreak or 
the averaged electrical field $\bar{E}$ (in the collisional case):  
\begin{align}
	&v_{\rm i,\,cf} \frac{\partial v_{\rm i,\,cf}}{\partial x} = \bar{E},\label{CollisionlessDimensionlessIonEquationOfMotion} \\
	&\nu  |v_{\rm i,\,c}| \, v_{\rm i,\,c} = \bar{E}. \label{CollisionalDimensionlessIonEquationOfMotion}
\end{align}
And, finally, focussed on the case of a single harmonic excitation,
\begin{align}
	&\tilde{j}(t) = -J \sin(t),\\
	&\tilde{Q}(t) = J \cos(t).
\end{align}

Lieberman's models were quite successful, and, to this date, they represent the standard \linebreak for any algebraic sheath model. They have, however, shortcomings
which directly arise from \linebreak the assumptions mentioned above:
\begin{itemize}
	  \item The restriction on single-frequency excitation is limiting. For technological reasons,  double and triple frequency plasmas 
		      have become frequently employed in the last years. Moreover, it is now known that even harmonically excited discharges may exhibit
					very unharmonics RF currents  \cite{Vandenplas1968,Klick96,Mussen06,Mussen07,Ziegler2008,CharlesBoswellPorteous1992}.
					
		\item Equally limiting is the concentration on the cases of purely collisional and purely collisionless ion motion.
		      Many technical plasmas are operated in the transition regime, where the mean free path and the sheath thickness are comparable 
					{\cite{gottscho92,Wu06,kim13}}. 
   
		\item The adoption of the Godyak step model implies that all ``thermal effects'' are neglected. 
		      This is especially serious when the excitation amplitude is small; in particular the phenomena of a finite sheath thickness and 
					a non-vanishing ``floating potential'' at zero modulation amplitude cannot be captured \cite{Brinkmann2007}.
\end{itemize}

These deficiencies are of very different nature. The assumption of a sinusoidal excitation \linebreak 
is merely a matter of convenience. The generalization to more complex current wave forms is easily 
possible (if cumbersome); several models which cover this point were published 
\cite{SheridanGorree1989,HorwitzPuzzer1990,RobicheBoyleTurnerEllingboe2003,JiangMaoWang2006,DewanMcNallyHerbert2001,
DewanMcNallyHerbert2002,TurnerChabert2012}.
\linebreak 
The restriction to the limit cases of collisionality is more severe; the derivations of Lieberman require 
explicite relations which express the ion density either in terms of the potential (collisionless case) or in terms of the field strength (collisional case). 
A direct generalization of the derivations to the transitional regime is not possible.

However, the most critical assumption is that of the step model.
Three recent publications have analyzed the situation in detail \cite{Brinkmann2007,Brinkmann2009,Brinkmann2011}. It was found that expression \eqref{StepModelField}
performs well in the electron depletion region but badly in the transition zone 
and the quasineutral zone where it misses the ambipolar field. 
This deficit causes a divergence of the ion density density at the sheath edge  but is of minor importance for the 
value of the sheath voltage. 
The error caused by \eqref{StepModelDensity}, however, is more critical: It cannot capture the residual
electron population in the depletion region which is needed to evaluate the conditions \eqref{DimensionlessCurrentVoltageRelation}
or $\eqref{DimensionlessFloatingRelation}$. 
Lieberman was thus forced to identify the electrode position $x_{\rm E}$ with the minumum value of the electron step $s_{\rm min}$. 
The analysis of  \cite{Brinkmann2007,Brinkmann2009,Brinkmann2011} found that the sheath voltage error caused by this uncertainty of the sheath shickness can be substantial.

\section{Employing the advanced algebraic approximation}

Publications \cite{Brinkmann2007,Brinkmann2009,Brinkmann2011} did not only analyze the problems caused by the Godyak step model but proposed also a cure. 
Based on a better approximate  -- for all practical purposes: exact --\linebreak  solution of the Boltzmann-Poisson problem 
\eqref{BoltzmannEquilibrium} and \eqref{PoissonEquation}, the so-called \textit{advanced algebraic approximation} (AAA)
was constructed. 
It is expressed in a new system of \textit{charge coordinates}, where the integration limit $\bar{s}$ was defined in \eqref{DimensionlessAverageCharge}:
\begin{align}
	q(x) = \int_{\bar{s}}^x n_{\rm i} (x^\prime) \, dx^\prime,
\end{align}
The approximation consists of expressions for the electrical field and the electron density,\linebreak all expressed in terms
of the difference of $q$ to the sheath charge $\tilde Q$, and the local values of the ion density and its derivative.
(The dependence on the dimensionless numbers $\vartheta$, $\nu$, and $\sigma$ is suppressed in the notation.)
The special functions $\Xi$ and $\Sigma$ which appear in \eqref{AAAE} and \eqref{AAAn} \linebreak are defined in terms of certain differential equations; 
 they are smooth and in fact analytical.  
We call them ``switch functions''; they switch the behavior of the field and density expressions
from electron depletion to quasineutrality (see Figs.~\ref{XiPlot} and \ref{SigmaPlot}). 
Finite values of $\vartheta$ lead to a thermal ``softening'' of the transition; the limit $\vartheta\to 0$ recovers the step model:
 \begin{align}
	E\left(q,n_{\rm i},n_{\rm i}^\prime,\tilde{Q}\right) &=  -\sqrt{  \vartheta  n_{\rm i}}\,\Xi_0\!\left(\frac{q-\tilde{Q} }{ \sqrt{\vartheta n_{\rm i}}}\right) - 
	 \vartheta {\partial n_{\rm i}\over\partial q}\,\Xi_1\!\left(\frac{q-\tilde{Q} }{ \sqrt{    \vartheta    n_{\rm i}}}\right),\label{AAAE}\\
	n_{\rm e}(q,n_{\rm i},n_{\rm i}^\prime,\tilde{Q}) &=  \Sigma_0\!\left(\frac{q-\tilde{Q} }{ \sqrt{    \vartheta    n_{\rm i}}}\right){n_{\rm i}}
         +  \Sigma_1\!\left(\frac{q-\tilde{Q} }{ \sqrt{    \vartheta    n_{\rm i}}}\right)
           \sqrt{  \vartheta  n_{\rm i}}\,{\partial n_{\rm i}\over\partial q}. \label{AAAn}
\end{align}
The AAA allows also an easy calculation of the phase-averages of the field and the density. \linebreak
It is useful to define a further set of functions $\Gamma$ and $N$, also termed ``switch functions'',
which depend functionally on $\tilde{Q}(t)$ but only locally on $q$ and $n_{\rm i}$:
\begin{align}
&\Gamma_0\left(q,n_{\rm i},\{\tilde{Q}\}\right) = 
{1\over 2\pi}\int_0^{2\pi} \Xi_0\!\left(\frac{q-\tilde{Q} }{ \sqrt{    \vartheta    n_{\rm i}}}\right)\, dt\,\sqrt{\vartheta n_{\rm i}}\,, \\[1.0ex]
&\Gamma_1\left(q,n_{\rm i},\{\tilde{Q}\}\right) = 
{1\over 2\pi}\int_0^{2\pi} \Xi_1\!\left(\frac{q-\tilde{Q} }{ \sqrt{    \vartheta    n_{\rm i}}}\right)\, dt, \\[1.0ex]
&N_0\left(q,n_{\rm i},\{\tilde{Q}\}\right) = 
 {1\over 2\pi}\int_0^{2\pi} \Sigma_0\!\left(\frac{q-\tilde{Q} }{ \sqrt{    \vartheta    n_{\rm i}}}\right)\, dt, \\[1.0ex]
&N_1\left(q,n_{\rm i},\{\tilde{Q}\}\right) = 
{1\over 2\pi}\int_0^{2\pi} \Sigma_1\!\left(\frac{q-\tilde{Q} }{ \sqrt{    \vartheta    n_{\rm i}}}\right)\, dt \, \frac{1}{\sqrt{\vartheta n_{\rm i}}}\,.
\end{align}
In terms of these special functions, the phase averages of the electrical field and the electron density are
local functions of $q$ and $n_{\rm i}(q)$ and linear forms in the derivative:
\begin{align}
&\bar{E}\left(q,n_{\rm i},n_{\rm i}^\prime,\{\tilde{Q}\}\right) =  -\Gamma_0\left(q,n_{\rm i},\{\tilde{Q}\}\right) -
\vartheta {\partial n_{\rm i}\over\partial q}\,\Gamma_1\left(q,n_{\rm i},\{\tilde{Q}\}\right),\label{PhaseAveragedField}\\[0.5ex]
&{\bar n}_{\rm e}\left(q,n_{\rm i},n_{\rm i}^\prime,\{\tilde{Q}\}\right)  =  N_0\left(q,n_{\rm i},\{\tilde{Q}\}\right) n_{\rm i}
 +\vartheta n_{\rm i} {\partial n_{\rm i}\over\partial q}   \,N_1\left(q,n_{\rm i},\{\tilde{Q}\}\right).
\end{align}

The ion model can also be  written in charge coordinates. Using the equation of continuity to express the velocity 
in terms of the density, the equation of motion reads 
\begin{align}
- \frac{1}{  n_{\rm i}^2}   \frac{\partial n_{\rm i}}{\partial q}  =  \bar{E}\left(q,n_{\rm i},n_{\rm i}^\prime,\{\tilde{Q}\}\right) + \frac{\nu}{\,n_{\rm i}^2}. 
\end{align}
Inserting the field expression and sorting the derivatives finally results in the sheath equation,
a quasi-linear differential equation of first order for ion density $n_{\rm i}(q)$:
\begin{align}
\left(- \frac{1}{n_{\rm i}^2}   +\vartheta \,\Gamma_1\left(q,n_{\rm i},\{\tilde{Q}\}\right)\right) \frac{\partial n_{\rm i}}{\partial q}   = 
-  \Gamma_0\left(q,n_{\rm i},\{\tilde{Q}\}\right)  + \frac{\nu }{n_{\rm i}^2}.  
\label{SheathEquation}
\end{align}

The properties of the sheath equation were discussed in \cite{Brinkmann2011}. Its most important feature is its inner singularity 
at a point $(q^*, n_{\rm i}^*)$ where the LHS and the RHS simultaneously vanish. \linebreak
This singularity -- which was identified as a ``collisionally modified Bohm point'' -- acts as an \linebreak inner boundary condition and
fixes the remaining degree of freedom of the sheath equation.\linebreak
The resulting ion density $n_{\rm i} = n_{\rm i}\bigl(q,\{\tilde{Q}\}\bigr)$ is a function of $q$ and a functional of the curve $\tilde{Q}$. \linebreak
Subsequently, the electrode position can be found from  the DC current-voltage curve;
\begin{align}
	\bar{j}   =   \sigma  \sqrt{\vartheta} \, \left(N_0\left(q,n_{\rm i},\{\tilde{Q}\}\right) n_{\rm i}
 +\vartheta n_{\rm i} {\partial n_{\rm i}\over\partial q}   \,N_1\left(q,n_{\rm i},\{\tilde{Q}\}\right)\right)\Bigl\vert_{q_{\rm E}}  - 1; \label{AAACurrentCurve}
\end{align}
under floating conditions one has to solve:
\begin{align}
	 \sigma  \sqrt{\vartheta}   \,\left(N_0\left(q,n_{\rm i},\{\tilde{Q}\}\right) n_{\rm i}
 +\vartheta n_{\rm i} {\partial n_{\rm i}\over\partial q}   \,N_1\left(q,n_{\rm i},\{\tilde{Q}\}\right)\right)\Bigl\vert_{q_{\rm E}}  =  1.
\label{AAAFloatingPotential}
\end{align}
The average sheath charge is the negative of that value,
\begin{align}
		\bar{Q} = \int_{q_{\rm E}}^{0} e n_{\rm i} \, \frac{1}{e n_{\rm i}}  dq = - q_{\rm E}. 
		\label{AverageChargeq}
\end{align}
Finally, the electrical potential across the sheath is calculated as
\begin{align}
V_{\rm sh}\left(\tilde{Q},\bar{Q},\{\tilde{Q}\}\right) &= 	-\int_{q_{\rm E}}^{q_{\rm B}}E\left(q,\tilde{Q}\right)\,\frac{1}{n_{\rm i}} dq  \\
              &= \int_{q_{\rm E}}^{q_{\rm B}}\left(  \sqrt{  \vartheta  n_{\rm i}}\,\Xi_0\!\left(\frac{q-\tilde{Q} }{ \sqrt{    \vartheta    n_{\rm i}}}\right) + 
	 \vartheta {\partial n_{\rm i}\over\partial q}\,\Xi_1\!\left(\frac{q-\tilde{Q} }{ \sqrt{    \vartheta    n_{\rm i}}}\right)\right) \, \frac{1}{n_{\rm i}}  dq.
	\nonumber
\end{align}
As the last step, one can construct transformation back into physical coordinates, 
\begin{align}
	x(q) = \bar s + \int_0^q   \frac{1}{n_{\rm i}}  dq.
\end{align}

\section{The algebraic sheath model}

For all practical purposes, the formulas given in the last chapter amount to an exact solution of the ``standard sheath model''.
Numerical studies confirmed that the deviation from the exact solution is in the percentage range (densities) or below (sheath voltage) 
\cite{Brinkmann2009}. \linebreak
However, the model is not yet algebraic, one must still solve the differential equation \eqref{SheathEquation}.
 In this section, we will employ further approximations to achieve an algebraic form.

We start by recalling that the presence  of the ambi\-polar field in \eqref{PhaseAveragedField} 
enables the solution to cross the critical point and enter the bulk regime.  Neglecting this contribution confines the model 
solely to the sheath, but the resulting error in the sheath voltage is only of order $\vartheta$.\linebreak We may safely neglect also 
other thermal effects in
the represen\-tation of the sheath field. The resulting model is exactly the step approximation:
\begin{equation}
E^{\rm (step)}\left(q,\tilde{Q}\right)  = \left\{
\begin{array}{c@{\quad:\quad}l}
           q-\tilde{Q}  & q < \tilde{Q}, \\
            0 & q \ge \tilde{Q}.
          \end{array}
\label{qStepModel}
\right.
\end{equation}
The phase average of $E^{\rm (step)}$ yields the step function version of the switch function $\Gamma_0$,
\begin{align}
	\bar{E}^{\rm (step)}\left(q,\{\tilde{Q}\}\right) = \frac{1}{2\pi} \int_0^{2\pi} E^{\rm (step)}\left(q,\tilde{Q}(t)\right) dt =
	-  	\Gamma^{\rm (step)}_0\left(q,\{\tilde{Q}\}\right). 
	\label{GammaStep}
\end{align}
The sheath differential equation now assumes a much simpler form. 
It is advantageous to express it in terms of the ion velocity, and to set -- as the now necessary boundary condition -- \linebreak 
the ion velocity at $\tilde{Q}_{\rm max}$ equal to the collisionally modified Bohm velocity \cite{Brinkmann2011}:
\begin{align}
& \frac{\partial v_{\rm i}}{\partial q}   =
   \Gamma^{\rm (step)}_0\left(q,\{\tilde{Q}\}\right)  - \nu v_{\rm i}^2, \label{GeneralIonModel}\\
 &v_{\rm i}(\tilde{Q}_{\rm max})	= v_{\rm B}(\nu).
\end{align}
The step function approximation can also be used for the sheath voltage, leading to
\begin{align}
V_{\rm sh}(\tilde{Q}) =\int_{q_{\rm E}}^{\tilde{Q}} 
						 (\tilde{Q}-q) \, \frac{1}{n_{\rm i}}  dq  =   \int_{q_{\rm E}}^{\tilde{Q}} 
						 (\tilde{Q}-q) \, |v_{\rm i}|  dq.
\end{align}
Unfortunately, a similar approximation for the current conditions \eqref{AAACurrentCurve} or \eqref{AAAFloatingPotential} is not feasible.
Simply neglecting thermal effects -- taking $\vartheta\to 0$ -- would render the conditions meaningless; 
any approximate evaluation would be instable due to the exponential nature of the relations. \linebreak
We thus keep for the moment the conditions \eqref{AAACurrentCurve} or \eqref{AAAFloatingPotential}, and will look later for other 
possibilities to determine the location of the electrode position $q_{\rm E} = -\bar{Q}$.

In the form displayed above, the ion  velocity $v_{\rm i}(q)$ or density $n_{\rm i}(q)$ are still given in terms of the solution of a nonlinear 
differential equation.  We first focus on the two limiting cases. The collision-free limit equation is as follows, with the boundary condition now
derived from the conventional Bohm condition
\begin{align}
&\frac{\partial v_{\rm i,\,cf}}{\partial q}   =  \Gamma^{\rm (step)}_0\left(q,\{\tilde{Q}\}\right) , \\
 &v_{\rm i,\,cf}\left(\tilde{Q}_{\rm max}\right)	= -\sqrt{\vartheta}.
\end{align}
We define the function $\Pi_0^{\rm (step)}(q)$ as the negative integral of $\Gamma_0^{\rm (step)}(q)$, 
\begin{align}
	  \Pi^{\rm (step)}_0\left(q,\{\tilde{Q}\}\right) = -\int_{\tilde{Q}_{\rm max}}^q\Gamma^{\rm (step)}_0\left(q^\prime,\{\tilde{Q}\}\right)\, dq^\prime,
		\label{PiStep}
\end{align}
and get thus the algebraic solution  
\begin{align}
	 v_{\rm i,\,cf}\left(q,\{\tilde{Q}\}\right) =  -\Pi^{\rm (step)}_0\left(q,\{\tilde{Q}\}\right)-\sqrt{\vartheta}.
\end{align}
The collisional limit is also algebraic and can be solved directly,
\begin{align}
 v_{\rm i,\,c}\left(q,\{\tilde{Q}\}\right) =- \sqrt{\frac{1}{\nu}  \Gamma^{\rm (step)}_0\left(q,\{\tilde{Q}\}\right)}. 
\end{align}
The general case of the ion model \eqref{GeneralIonModel} does not have a similarly simple algebraic solution
but its physical content is transparent: The electrical force is balanced by inertia and friction, and thus dominated by
whatever is larger. Numerical experiments have shown that it is reasonable to employ the approximation of a quadratic harmonic mean,
\begin{align}
	v_{\rm i}\left(q,\{\tilde{Q}\}\right) =
	-\displaystyle\left(\frac{1} {\left(\Pi^{\rm (step)}_0\left(q,\{\tilde{Q}\}\right)+\sqrt{\vartheta}\right)^2}  +
		\frac{\nu  }{\Gamma^{\rm (step)}_0\left(q,\{\tilde{Q}\}\right)}\right)^{-\frac{1}{2}}
		\label{QuadraticHarmonicMean}
\end{align}

Together, the described equations constitute an algebraic (= ``closed form'') model to calculate the time-resolved sheath voltage $V_{\rm sh}$.
However, it is not yet very convenient to use, owing to the presence of the functions  $\Gamma^{\rm (step)}_0\left(q,\{\tilde{Q}\}\right)$
and  $\Pi^{\rm (step)}_0\left(q,\{\tilde{Q}\}\right)$ which depend not only on the variable $q$ but also
functionally on the modulation $\tilde{Q}$. We therefore implement a sequence of additional approximations which replace this dependence
with a dependence on the modulation amplitude defined in \eqref{EffectiveModulationAmplitude} which reads in dimensionless units 
\begin{align}
\Delta Q = \sqrt{\frac{1}{2\pi} \int_0^{2\pi} 	{\tilde Q}(t)^2\, dt}. \label{DeltaQNorm}
\end{align}

First, consider the function $\Gamma^{\rm (step)}_0\bigl(q,\{\tilde{Q}\}\bigr)$. It functionally depends on the charge $\tilde{Q}(t)$, \linebreak but all curves yield the same qualitative behavior.
It is $-q$ for $q<\tilde{Q}_{\rm min}$,   zero for $q>\tilde{Q}_{\rm max}$, \linebreak and monotonically
decreasing with positive curvature in between. We replace this family of functions by a
qualitatively similar but mathematically simpler model which exhibits the same asymptotic behavior
as the originals for $|q|\ge\sqrt{3}\Delta Q$ and matches the branches smoothly with an interpolating 
parabola of positive curvature:
\begin{align}
	\Gamma(q,\Delta Q) &=  
	\left\{
	\begin{array}{c@{\quad:\quad}l}
							-q  & q < - \sqrt{3}\Delta Q,\\[0.0ex]
						\displaystyle -\frac{\left(\sqrt{3}\Delta Q-q\right)^2}{4 \sqrt{3} \Delta Q} & |q| \leq \sqrt{3}\Delta Q,\\[1.0ex]
							0 & q> \sqrt{3}\Delta Q.  
	\end{array}
	\right.
	\label{GammaModel}
	\end{align}
The choice of matching points minimizes the deviation to the original family of functions, in the sense that
the integral over the difference vanishes,
\begin{align}
	  \int_{-\infty}^\infty \Gamma^{\rm (step)}_0\left(q,\{\tilde{Q}\}\right) - \Gamma(q,\Delta Q) \, dq   = 0.
\end{align}

Numerical experiments have convinced us that form \eqref{GammaModel} is a reasonable approximation
of $\Gamma^{\rm (step)}_0\bigl(q,\{\tilde{Q}\}\bigr) $
for all ``generic''  charge modulation functions ${\tilde Q}(t)$. 
(To elaborate the point: The formula is exact when ${\tilde Q}(t)$ is a saw tooth; it is generally very satisfactory when the fundamental in
${\tilde Q}(t)$ is dominant. It is less appropriate for square wave or pulsed modulation; in this case the development should be
carried out with the original formulas \eqref{GammaStep} and \eqref{PiStep}.)
Once adopted, approximation \eqref{GammaModel} can be used to define 
\begin{align}
\Pi(q,\Delta Q) &=  
\left\{
\begin{array}{c@{\quad:\quad}l}
          \displaystyle \frac{\Delta Q^2}{2}+\frac{q^2}{2}  & q < - \sqrt{3}\Delta Q,\\[0.0ex]
				   \displaystyle\frac{\left(\sqrt{3} \Delta Q-q\right)^3}{12 \sqrt{3}\Delta Q}  & |q| \leq \sqrt{3}\Delta Q,\\[1.0ex]
            0 & q> \sqrt{3}\Delta Q.  
\end{array}
\right.
\label{PiModel}
\end{align}
The ion velocity $v_{\rm i}(q,\Delta Q)$ and the ion density $n_{\rm i}(q,\Delta Q)$ are then 
\begin{align}
	&v_{\rm i}(q,\Delta Q) = 
	-\displaystyle\left(\frac{1}{\bigl(\Pi(q,\Delta Q)+\sqrt{\vartheta}\bigr)^2}  +
		\frac{\nu  }{\Gamma(q,\Delta Q)}\right)^{-\frac{1}{2}},\label{IonVelocityModel}\\
	&n_{\rm i}(q,\Delta Q) = 
	\displaystyle\left(\frac{1} {\bigl(\Pi(q,\Delta Q)+\sqrt{\vartheta}\bigr)^2}  +
		\frac{\nu  }{\Gamma(q,\Delta Q)}\right)^{\frac{1}{2}},
\end{align}
and the charge-voltage relation of the sheath is:
\begin{align}
V_{\rm sh}(\tilde{Q},q_{\rm E},\Delta Q) =\int_{q_{\rm E}}^{\tilde{Q}} 
						  \frac{\tilde{Q}-q}{n_{\rm i}(q,\Delta Q)}\, dq  =   \int_{q_{\rm E}}^{\tilde{Q}} 
						 (\tilde{Q}-q) \, |v_{\rm i}(q,\Delta Q)|  dq.
						\label{NearlyFinalFormula}
\end{align}

Formula \eqref{NearlyFinalFormula} is nearly the desired result, an algebraic description of the sheath behavior.
Unfortunately, however, the integral cannot be carried out analytically. We therefore resort to an approximate evaluation 
in the spirit of Kepler's barrel rule: 
Namely, we interpolate the function $|v_{\rm i}| $ by a quadratic parabola, with nodes specified at $q_{\rm E}$, $(q_{\rm E}+\tilde{Q})/2$,
and $\tilde{Q}$. Inserting this parabola, evaluating the integral, and substituting $\tilde{Q}=Q-\bar{Q}$ and $q_{\rm E}= -\bar{Q}$ finally leads 
to the desired algebraic formula for the sheath voltage
\begin{align}
	V_{\rm sh}(Q,\bar{Q},\Delta Q) =
	\frac{1}{6} \bigl( |v_{\rm i}(-\bar{Q},\Delta Q)|+ 2  |v_{\rm i}(-\bar{Q}+\textstyle\frac{1}{2}Q ,\Delta Q)| \bigr)
	Q^2.
\end{align}

It may be helpful to view our results in dimensional units. We define the quantity $u$ as the absolute value of
the ion velocity $v_{\rm i}$ and obtain the following, where $q$ and $\Delta Q$ have units $\rm{As}/{m}^2$
and $\Pi(q,\Delta Q)$ and $\Gamma(q,\Delta Q)$ are litterally unchanged from above: 
\begin{align}
	u(q,\Delta Q) = 
	\left(\frac{1}{\bigl(\Pi(q,\Delta Q)/\epsilon_0 \Psi_{\rm i} +\sqrt{T_{\rm e}/m_{\rm i}}\bigr)^2}  +  \frac{\pi}{2}\frac{m_{\rm i}}	{e \lambda_{\rm i}}
		\frac{\epsilon_0}{\Gamma(q,\Delta Q)}\right)^{-\frac{1}{2}}.
\end{align}
The inverse of this quantity, times the ion flux, is the ion density
\begin{align}
	n_{\rm i}(q,\Delta Q) = 
	\Psi_{\rm i}\left(\frac{1}{\bigl(\Pi(q,\Delta Q)/\epsilon_0 \Psi_{\rm i} +\sqrt{T_{\rm e}/m_{\rm i}}\bigr)^2}  +  \frac{\pi}{2}\frac{m_{\rm i}}	{e \lambda_{\rm i}}
		\frac{\epsilon_0}{\Gamma(q,\Delta Q)}\right)^{\frac{1}{2}}.
\end{align}
These functions allow to explicitly express the DC current characteristics of the sheath. 
Employing formula \eqref{AAACurrentCurve} in leading order in $\vartheta$, we get
\begin{align}
	\bar{j}   = e   s_{\rm e}  \sqrt{\frac{ T_{\rm e}  }{2\pi m_{\rm e}}}   {1\over T}\int_0^{T} 
	\Sigma_0\!\left(\frac{q_{\rm E}-\tilde{Q}(t) }{\sqrt{\epsilon_0 T_{\rm e} n_{\rm i}(q_{\rm E})}}\right) dt \, n_{\rm i}(q_{\rm E}) -  e\Psi_{\rm i} . \label{AAACurrentCurvePhysical}
\end{align}
For a practical evaluation of this expression it may be useful to know that the switch function $\Sigma_0(\xi)$ can be approximated as
\begin{align}
	\Sigma_0(\xi) \approx \exp\left({-\frac{{\xi ^2}/{2}+1}{0.685\, \exp({2 \xi })+1}}\right).
\end{align}
Once the charge coordinate position $q_{\rm E}$ of the electrode is known, for instance by imposing the floating condition $\bar{j}=0$,
one can set the average sheath charge $\bar{Q}=-q_{\rm E}$ and finally calculate the desired charge-voltage relation from
\begin{align}
	V_{\rm sh}\bigl(Q,\bar{Q},\Delta Q\bigr) = \frac{1}{6 e \epsilon_0 \Psi_{\rm i}}
 \left( u(-\bar{Q},\Delta Q)+ 2  u(-\bar{Q}+\textstyle\frac{1}{2}Q ,\Delta Q) \right) 	Q^2. \label{VQ}
\end{align}

\section{Example and comparison with particle in cell}

We now compare our algebraic charge-voltage relation \eqref{VQ} with the predictions of a more 
fundamental modeling approach, namely a fully self-consistent particle-in-cell simulation.
PIC is truly kinetic, i.e., makes no priori assumptions on the particle distribution functions, \linebreak
and is therefore particularly suited for an investigation of the boundary sheath where both ions and electrons are 
far from equilibrium. We employ \textit{yapic}, an explicite 1d3v version   of the PIC algorithm which is described
in \cite{TurnerDerzsiDonkoEreminKellyLafleurMussenbrock2012,TrieschmannShihabSzeremleyElgendyGallianEreminBrinkmannMussenbrock2013}. 
The discharge gas is argon, at $T_{\rm N}=300\,\rm{K}$;\linebreak the pressure values are $p=0.1\,{\rm Pa}$ (nearly collisionless regime),
$p=1\,{\rm Pa}$ (transition regime), and $p=10\,{\rm Pa}$ (collisional regime). For all pressures values, both a single frequency and a double frequency 
excitation are studied See table 1 for details of the simulation parameters. \linebreak In all simulation runs, the net DC current is assumed to be zero, 
i.e., the boundary sheaths are driven under floating conditions.

The PIC simulation gives access to all relevant quantities. Figs.~\ref{PIC1FResults} (one frequency) and \ref{PIC2FResults} (two frequencies) 
show the ion density $n_{\rm i}(x)$ and the phase-averaged electron density $\bar{n}_{\rm e}(x)$.\linebreak
The charge-voltage relations -- figs.~\ref{QVResults1F}  and \ref{QVResults2F} --  can be obtained from the monitoring the electrical field at the 
electrode and the integral of the field from the electrode to the sheath edge. \linebreak
Further quantities taken from the PIC simulations are the mean ion flux $\Psi_{\rm i}$ at the electrode, 
the phase-averaged sheath charge $\bar{Q}$, and the electron temperature $T_{\rm e}$. 

Based on these quantities, the charge-voltage characteristics of our algebraic model are calculated and plotted into the same
figures \ref{QVResults1F}  and \ref{QVResults2F}. We apply two different procedures. \linebreak
In the first one (dashed), we use all quantities of the PIC simulation; in the second (dotted), we do not utilize the PIC sheath charge $\bar{Q}$
but evaluate the floating condition \eqref{AAACurrentCurvePhysical}.

The agreement is excellent. When all information from PIC is used, the characteristics of the collision-less and the
collisional case are nearly exactly reproduced. The agreement is less \linebreak  spectacular in the transition regime; our interpolation 
\eqref{QuadraticHarmonicMean}
is only a rough representation of the complicated ion dynamics in this regime. When the PIC information in $\bar{Q}$ is not used, 
the agreement is only slightly worse; except for the one frequency/ $0.1\,{\rm Pa}$ case where the calculation of the average sheath charge  
(or electrode position) shows a considerable offset.
The same conclusion is also suggested by the phase-resolved voltages of figs.~\ref{Voltage0p1Pa}, \ref{Voltage1Pa}, and \ref{Voltage10Pa}.
A closer inspection of the PIC data reveals that the deviation at 1f/$0.1\,{\rm Pa}$ is caused by a 
highly non-Maxwellian electron energy distribution due to stochastic heating.

\pagebreak
\section{Summary and discussion}

In this manuscript, we have presented a novel algebraic model for the electrical behavior of RF modulated plasma boundary sheaths. 
Our investigation was motivated by a critical \linebreak  assessment of the pioneering Lieberman models  \cite{Lieberman1988,Lieberman1989}
and shared many of their assumptions: \linebreak The focus on the RF regime, where the applied frequency lies between the plasma frequencies \linebreak
 of the ions and the electrons, 
$\omega_{\rm pi} \ll \omega_{\rm RF} \ll \omega_{\rm pe}$,  the concentration on only one species of singly charged positive ions 
with ``chemistry'' (ionization) neglected, 
and the assumption of a one-dimensional Cartesian geometry.
We have corrected, however, the three fundamental weaknesses of the Lieberman models,
namely their limitations to a single driving frequency, \linebreak to the regime of large applied voltages 
(compared to the thermal voltage $T_{\rm e}/e$), and to the two cases of either highly collisional or completely 
collision-free motion. 

Our new algebraic sheath model captures the plasma sheath dynamics for a wide range of frequencies, waveforms, amplitudes, 
and collisionality. A comparison with self-consistent particle-in-cell simulations has demonstrated the excellent accuracy of our final expression. 
We believe that our model will find many useful approcations in the future.

\section{Acknowledgments}

The authors gratefully acknowledge support by the  Deutsche Forschungsgemeinschaft via SFB-TR87 and the Ruhr-University Research School.

\pagebreak

\pagebreak

\pagebreak

 \begin{figure}[h!]
  \centering
 \includegraphics[width=1.0\textwidth]{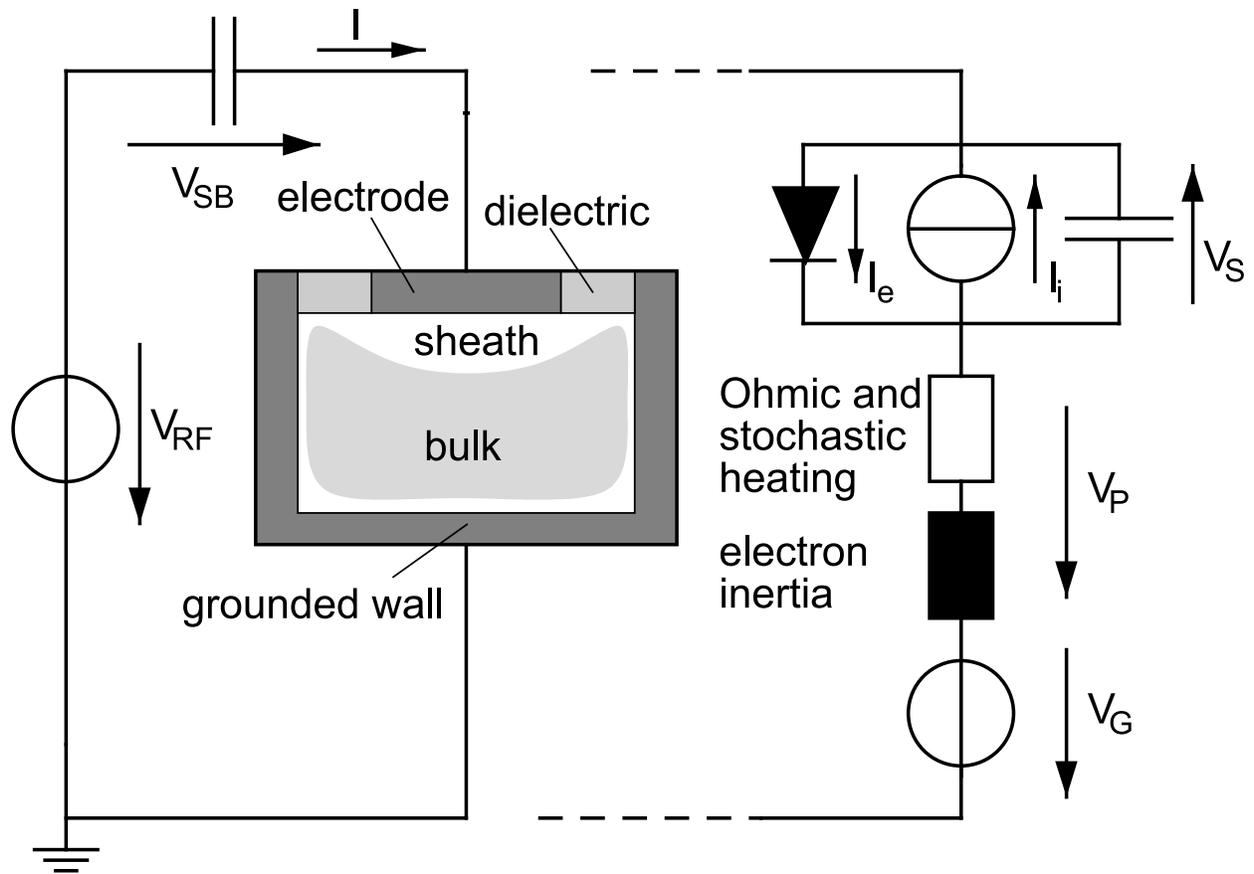}
 \caption{Lumped element equivalent circuit (global model) of a capacitively coupled plasma (CCP).
          The three element subcircuit of a diode, a current source, and a nonlinear 
					capacitor represents the electrical behavior of the plasma boundary sheath.}\label{EquivalentCircuit}
 \end{figure}

\pagebreak

 \begin{figure}[h!]
  \centering
 \includegraphics[width=1.0\textwidth]{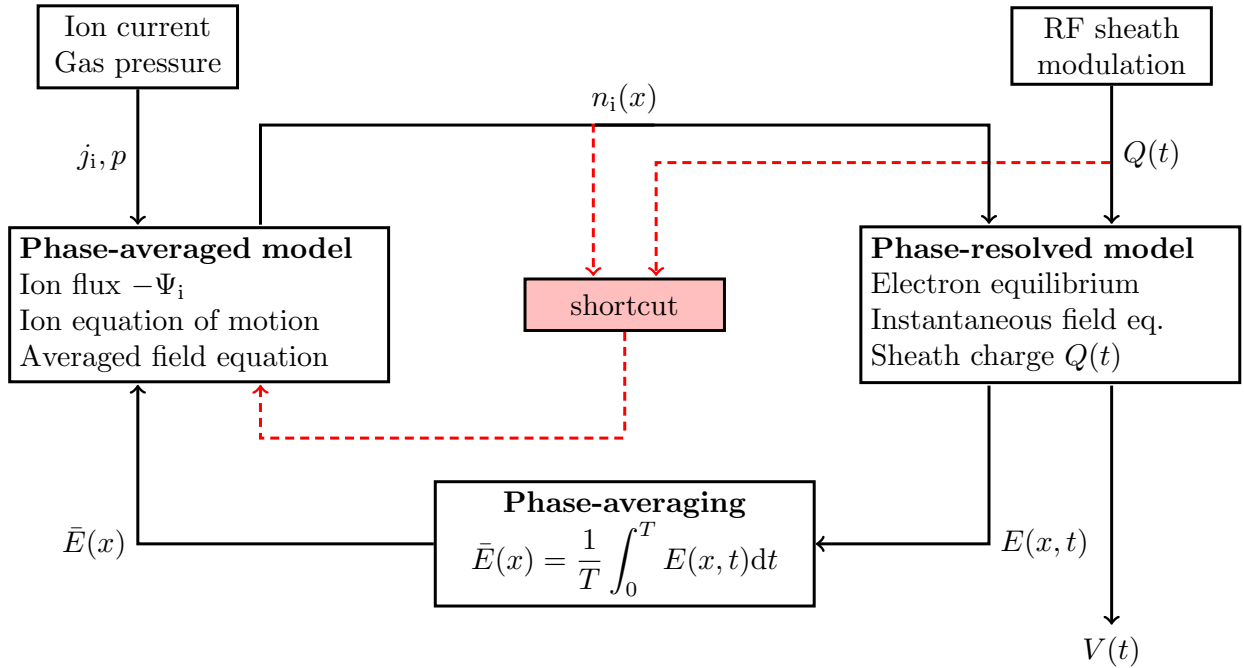}
 \caption{Schematic depiction of the problem posed by the ``standard'' plasma sheath model and of the mathematical shortcut which is required 
to solve it algebraically.}\label{Shortcut}
 \end{figure}

\pagebreak

 \begin{figure}[h!]
  \centering
 \includegraphics[width=0.8\textwidth]{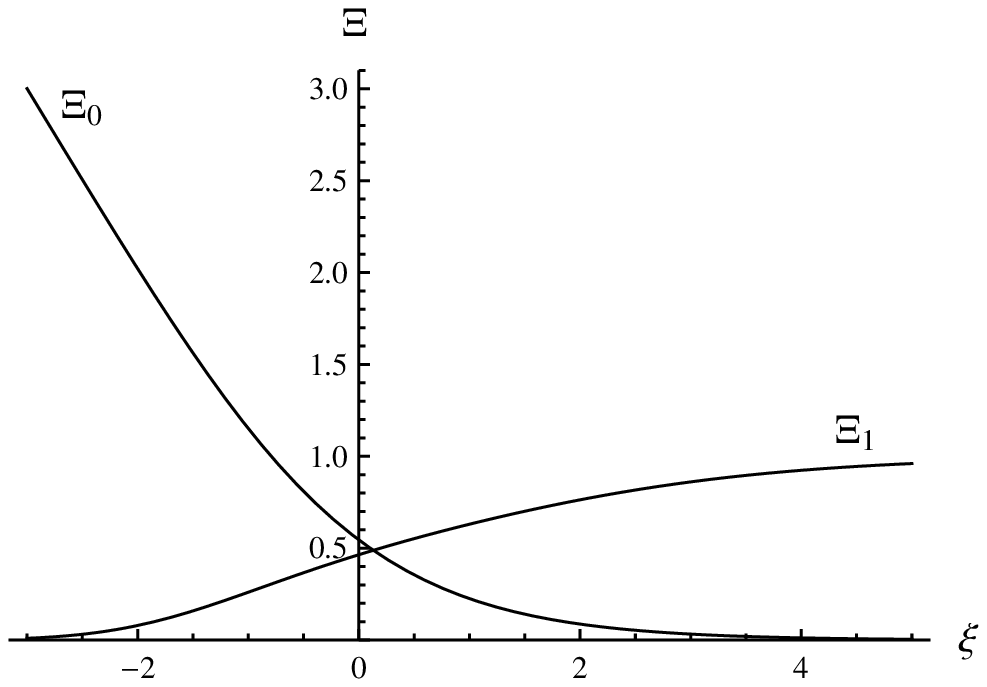}
\caption{The switch functions $\Xi_0(\xi)$ and $\Xi_1(\xi)$ in dependence of their argument. 
         The limit $\xi\ll 0$ describes the depletion region, the limit $\xi\gg 0$ the ambipolar region.}
\label{XiPlot}
 \end{figure}

\pagebreak

 \begin{figure}[h!]
  \centering
 \includegraphics[width=0.8\textwidth]{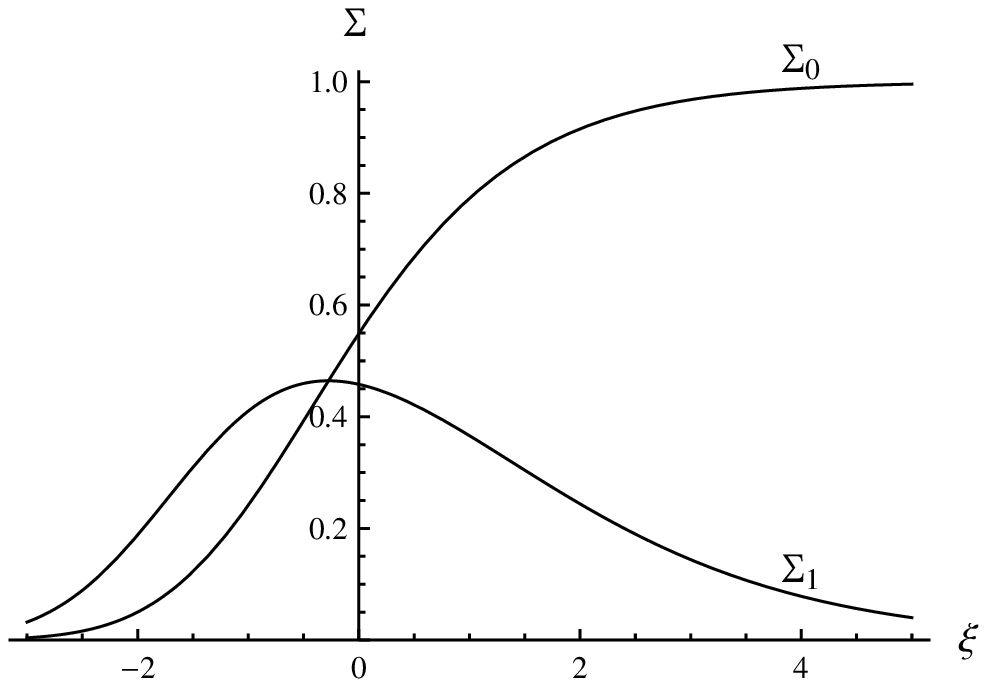}
\caption{The sheath functions $\Sigma_0(\xi)$ and $\Sigma_1(\xi)$ in dependence of their argument. 
         The limit $\xi\ll 0$ describes the depletion region, the limit $\xi\gg 0$ the ambipolar region.}
\label{SigmaPlot}
 \end{figure}

\pagebreak

\begin{table}[p]
\begin{center}
\begin{tabular}{l*{7}{c}r}
Case                 & 1f/0.1Pa  & 1f/1Pa   & 1f/10Pa    & 1f/0.1Pa   & 1f/1Pa   & 2f/10Pa  & Units\\
\hline
Pressure             & 0.1       & 1        & 10         & 0.1        & 1        &  10      &   Pa\\
Mean free path       & 0.078     & 0.0078   & 0.00078    & 0.078      & 0.0078   &  0.00078 &   m\\ 
Electrode gap        & 0.1       & 0.05     & 0.05       & 0.01       &  0.05    & 0.05     &   m\\
Amplitude 13.56 MHz  & 400       & 200      & 200        & 200        & 100      & 100      &   V\\
Amplitude 27.12 MHz  & --        & --       & --         & 200        &  200     &  200     &   V\\
Ion flux             & 1.85      & 1.48     & 1.58       & 2.55       &  1.47    &  1.35    &   $10^{18}\,\rm{m}^{-2}\rm{s}^{-1}$\\
Electron temperature & 13        & 4.3      & 2.5        & 13         &  4.3     &  2.5     &  eV\\
$\bar{Q}$ from PIC   & 2.19      & 1.83     & 2.72       & 2.30       &  1.65    &  2.28    & $10^{-7}\, {\rm As}/{\rm m}^2$ \\
$\bar Q$ from floating condition    & 1.94      & 1.79     & 2.78       & 2.28       & 1.60     &  2.35    & $10^{-7}\, {\rm As}/{\rm m}^2$ \\
\end{tabular} \\[5.0ex]
\caption{Parameters of the discharge simulations with the PIC code \textit{yapic} and of the input parameters of the algebraic sheath model.}  
\end{center}
\end{table}

\pagebreak

 \begin{figure}[h!]
  \centering
 \includegraphics[width=1.0\textwidth]{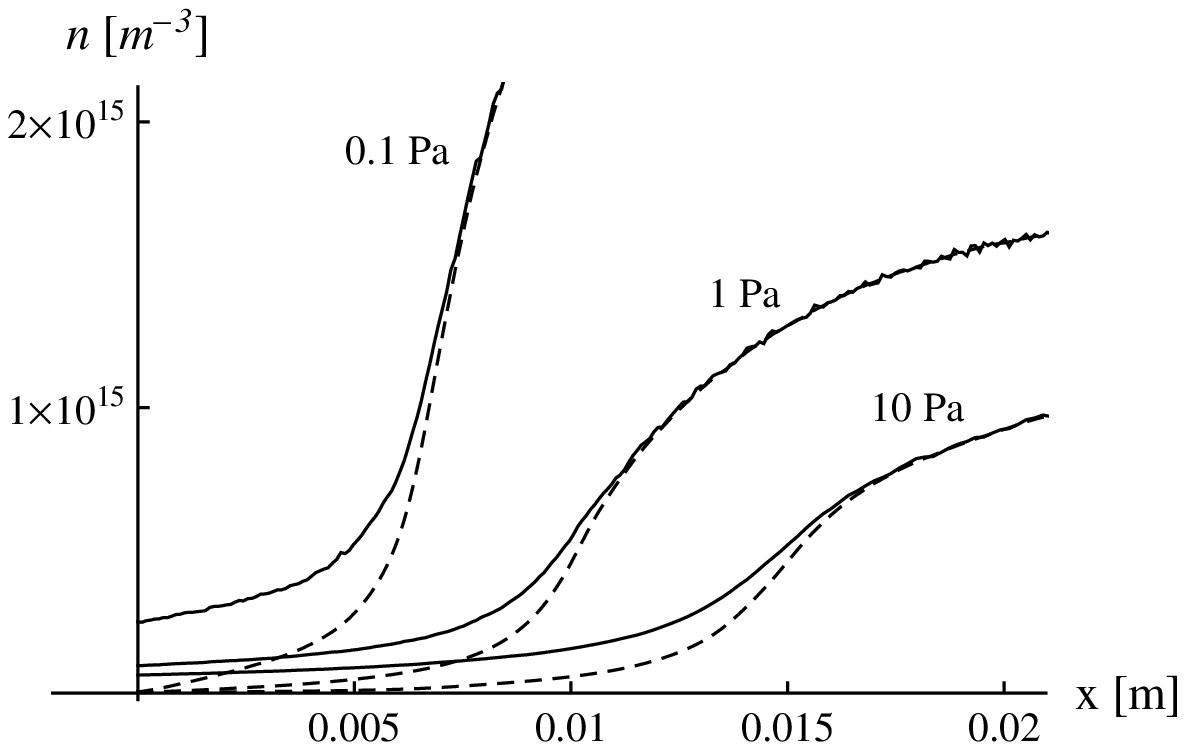}
\caption{Particle densities in the boundary sheath of a 1f-CCP, as obtained by the PIC code \textit{yapic}, for the
         pressure cases of 0.1\,Pa, 1\,Pa, and 10\,Pa. The ion density is solid, the phase-averaged electron density is dashed.}
\label{PIC1FResults}
 \end{figure}

\pagebreak

 \begin{figure}[h!]
  \centering
 \includegraphics[width=1.0\textwidth]{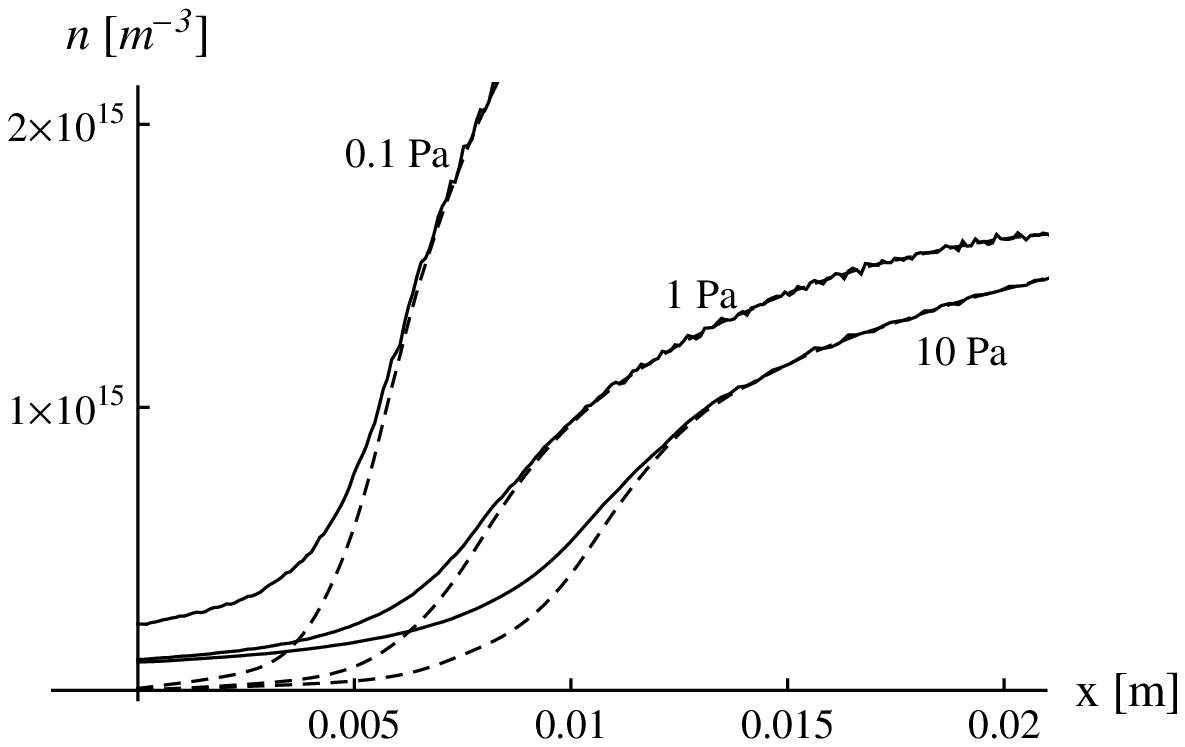}
\caption{Particle densities in the boundary sheath of a 2f-CCP, as obtained by the PIC code \textit{yapic}, for the
         pressure cases of 0.1\,Pa, 1\,Pa, and 10\,Pa. The ion density is solid, the phase-averaged electron density is dashed.}
\label{PIC2FResults}
 \end{figure}

\pagebreak

 \begin{figure}[h!]
  \centering
 \includegraphics[width=1.0\textwidth]{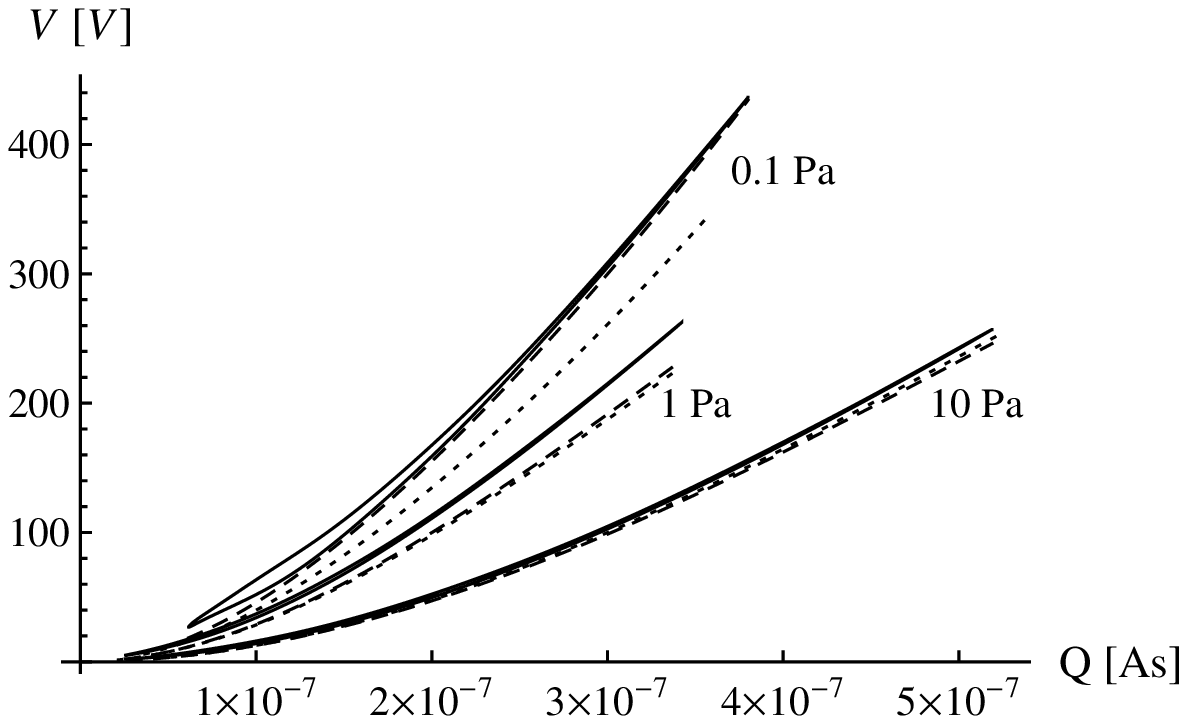}
\caption{Charge-voltage relations of the 1f-CCP, for the
         pressure cases of 0.1\,Pa, 1\,Pa, and 10\,Pa. The PIC results are solid, the results from the algebraic model with $\bar{Q}$ taken from PIC are dashed,
				  those obtained with the floating condition are dotted.}
\label{QVResults1F}
 \end{figure}

\pagebreak

 \begin{figure}[h!]
  \centering
 \includegraphics[width=1.0\textwidth]{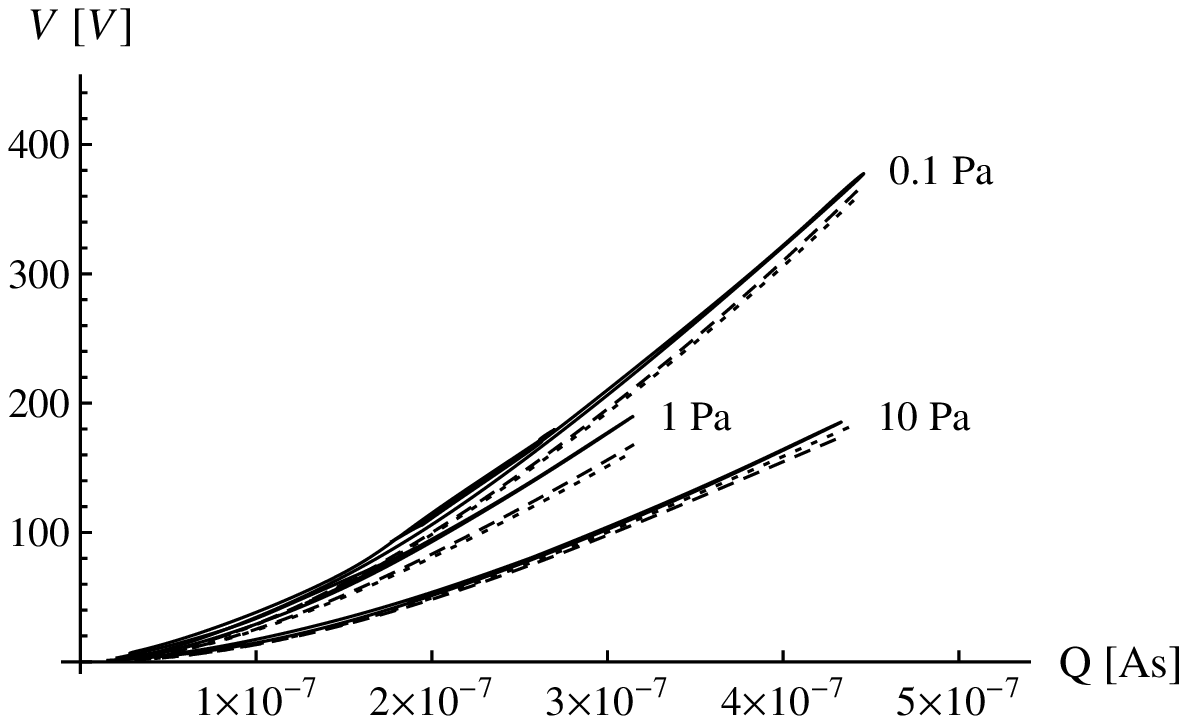}
\caption{Charge-voltage relations of the 2f-CCP, for the
         pressure cases of 0.1\,Pa, 1\,Pa, and 10\,Pa. The PIC results are solid, the results from the algebraic model with $\bar{Q}$ taken from PIC are dashed,
				  those obtained with the floating condition are dotted.}
\label{QVResults2F}
 \end{figure}

\pagebreak

 \begin{figure}[h!]
  \centering
 \includegraphics[width=1.0\textwidth]{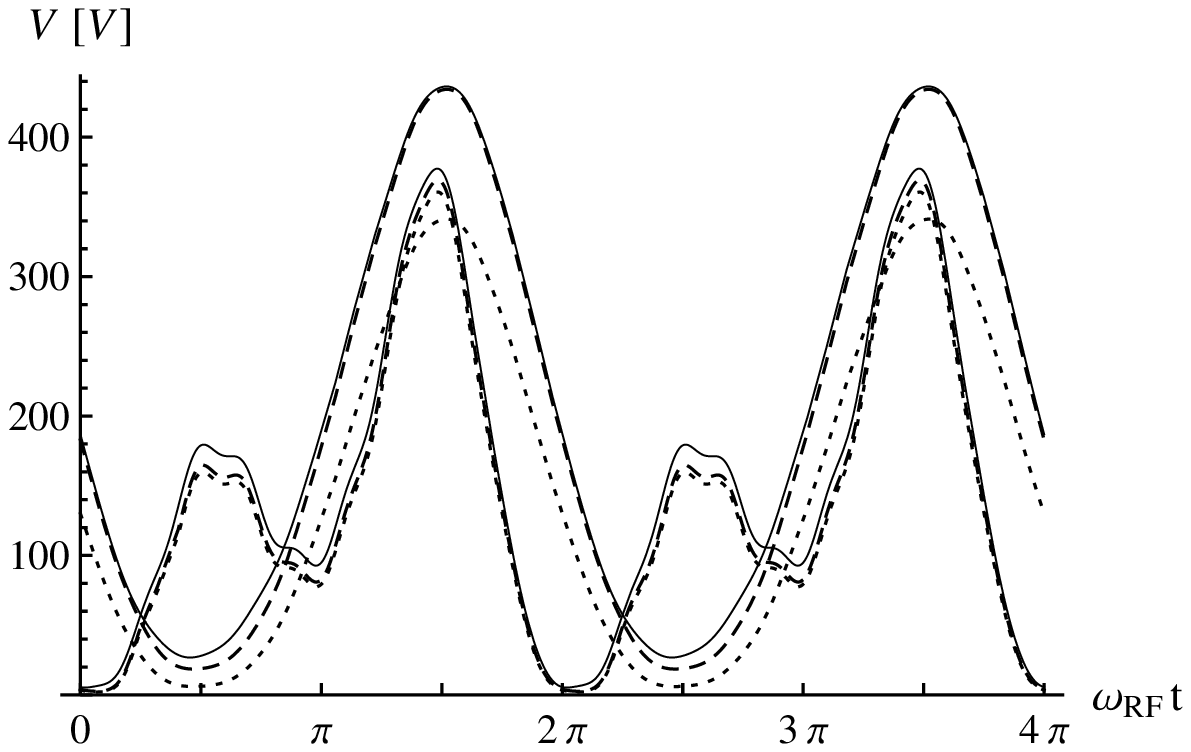}
\caption{Phase-resolved sheath voltages $V_{\rm sh}(t)$ for the 1f and 2f cases at a pressure $p = 0.1\,{\rm Pa}$. The PIC results are solid, the results from the algebraic model with $\bar{Q}$ taken from PIC are dashed,
				  those obtained with the floating condition are dotted.}
\label{Voltage0p1Pa}
 \end{figure}

\pagebreak

 \begin{figure}[h!]
  \centering
 \includegraphics[width=1.0\textwidth]{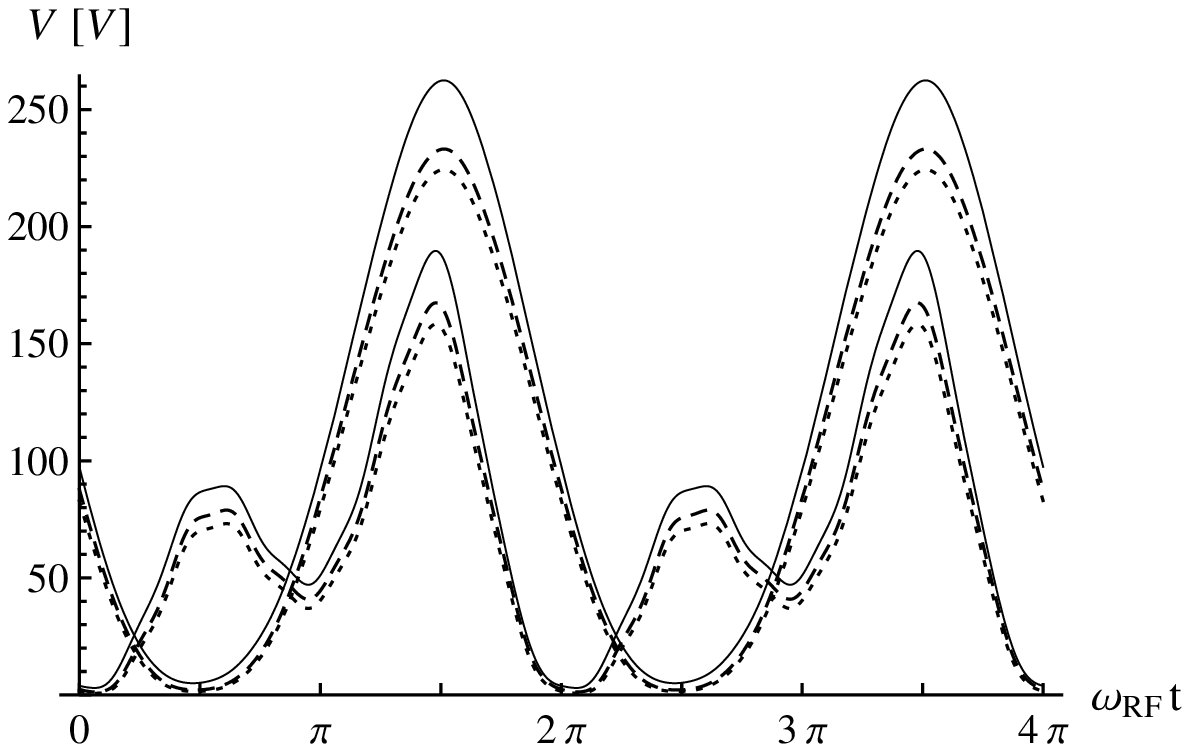}
\caption{Phase-resolved sheath voltages $V_{\rm sh}(t)$ for the 1f and 2f cases at a pressure $p = 1\,{\rm Pa}$. The PIC results are solid, the results from the algebraic model with $\bar{Q}$ taken from PIC are dashed,
				  those obtained with the floating condition are dotted.}
\label{Voltage1Pa}
 \end{figure}

\pagebreak

 \begin{figure}[h!]
  \centering
 \includegraphics[width=1.0\textwidth]{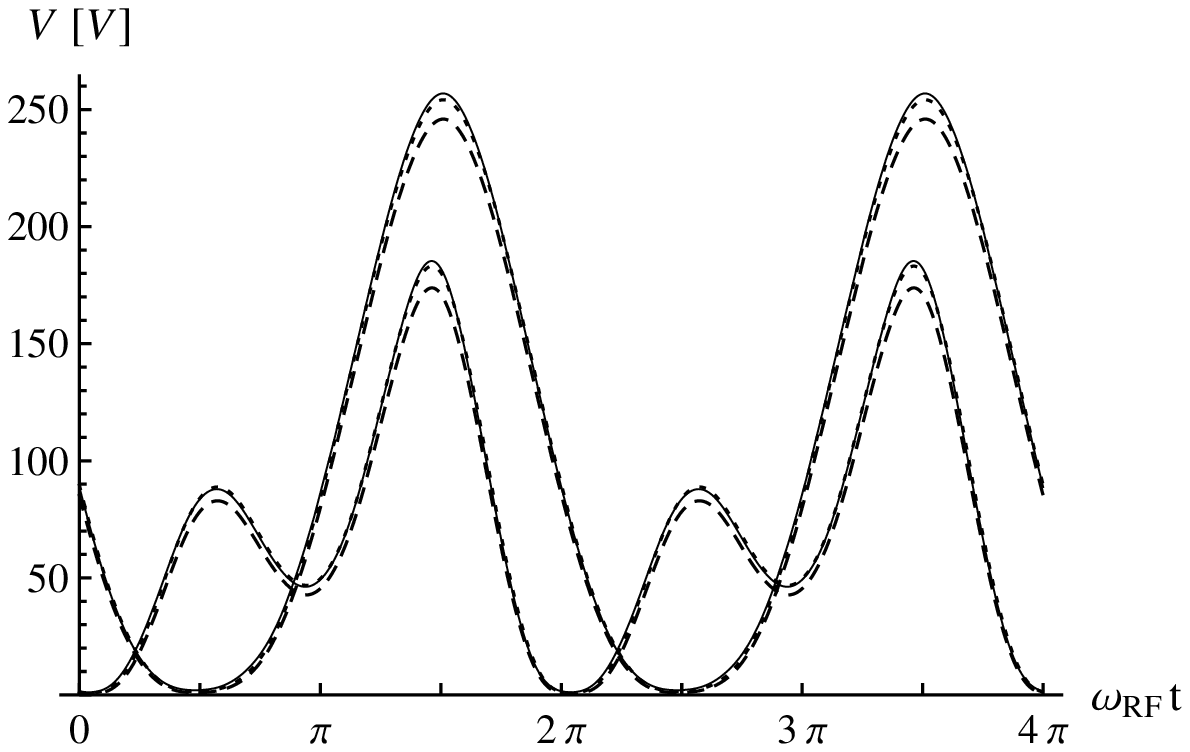}
\caption{Phase-resolved sheath voltages $V_{\rm sh}(t)$ for the 1f and 2f cases at a pressure $p = 10\,{\rm Pa}$. The PIC results are solid, the results from the algebraic model with $\bar{Q}$ taken from PIC are dashed,
				  those obtained with the floating condition are dotted.}
\label{Voltage10Pa}
 \end{figure}

\end{document}